\documentstyle[prl,aps,twocolumn,epsf]{revtex}
\begin{document}
\twocolumn[\hsize\textwidth\columnwidth\hsize\csname
@twocolumnfalse\endcsname

\draft

\title{Prediction of a New Type of Strain Induced Conduction Band
Minimum in Embedded Quantum Dots}

\author{A.~J.~Williamson and Alex~Zunger}
\address{National Renewable Energy Laboratory, Golden, CO 80401}
\author{A.~Canning}
\address{NERSC, Lawrence Berkeley National Laboratory, Berkeley, CA94720}

\date{\today}
\maketitle

\begin{abstract}
\begin{quote}\parbox{16 cm}{\small 
Free standing InP quantum dots have previously been theoretically and
experimentally shown to have a direct band gap across a large range of
experimentally accessible sizes.  We demonstrate that when these dots
are embedded coherently within a GaP barrier material, the effects of
quantum confinement in conjunction with coherent strain suggest there
will be a critical diameter of dot ($\approx$60\AA~), above which the
dot is direct, type I, and below which it is indirect, type II.
However, the strain in the system acts to produce another conduction
state with an even lower energy, in which electrons are localized in
small pockets at the {\em interface} between the InP dot and the GaP
barrier.  Since this conduction state is GaP $X_{1c}$-derived and the
highest occupied valence state is InP, $\Gamma$-derived, the
fundamental transition is predicted to be indirect in both real and
reciprocal space (``type II'') for all dot sizes.  This effect is
peculiar to the strained dot, and is absent in the free-standing dot.
} \end{quote}
\end{abstract}

]
\narrowtext 

\newpage

There are two leading techniques for fabricating InP quantum dots; (i)
colloidal growth, producing unstrained, chemically passivated
dots\cite{micic94} and (ii) dots grown by the Stranski-Krastanov
mechanism that are embedded within a semiconductor barrier such as
GaInP\cite{samuelson94,kurtenbach95}.  In a recent study of the
electronic structure of colloidally grown InP dots\cite{fu97} it was
found that these dots have a direct band gap at the $\Gamma$-point of
the Brillouin zone for all experimentally accessible sizes.  In
contrast, it has previously been shown that for free-standing GaAs
dots, the band gaps can undergo a transition from direct to indirect
as a result of quantum size effects\cite{alberto_apl}. The effects of
quantum confinement are that as one decreases the size of the dot, all
of the conduction levels are pushed up in energy at a rate reflecting
approximately the inverse of the electron effective mass.  Since the
$\Gamma_{1c}$ masses are generally lighter than $X_{1c}$
masses\cite{bornstein}, reduced sizes can transform a direct dot into
an indirect dot if the initial $\Gamma_{1c}$-$X_{1c}$ separation in
the bulk is not too large.  It has been predicted\cite{alberto94} that
in GaAs, where the bulk $\Gamma_{1c}$-$X_{1c}$ separation is only
0.55~eV\cite{bornstein}, a free-standing, zero-pressure dot will
become indirect at a dot size of 40\AA~, whereas an AlAs-embedded GaAs
dot will become indirect at 80\AA~\cite{alberto94}.  Since, however,
in InP the $\Gamma_{1c}$-$X_{1c}$ bulk separation is large
(0.94~eV\cite{bornstein}), calculations\cite{fu97} have predicted that
this separation is not overcome by quantum-size effects, and
free-standing dots will remain direct at all sizes at zero pressure.

In this paper we are interested in investigating whether InP quantum
dots embedded within a GaP matrix exhibit a direct-to-indirect
transition as a function of size and thereby exhibit significantly
different electronic structure to free standing InP quantum dots that
are direct for all sizes.  We find indeed that for spherical InP dots
smaller than 60\AA~in diameter, the $X_{1c}$ conduction band of the
unstrained GaP barrier is lower than the $\Gamma_{1c}$-state of the
InP.  This is analogous to a ``type II in real-space and in reciprocal
space'' state familiar\cite{ploog89} in AlAs/GaAs nanostructures.
Surprisingly, however, we find that under the influence of the
hydrostatic and biaxial strain present at the GaP/InP interface, a new
conduction state emerges that is lower in energy than both the
unstrained bulk GaP $X_{1c}$-state and the InP $\Gamma_{1c}$-state.
This qualitatively new type of state is localized at the interface of
the dot and its barrier, and is indirect ($X$-like).  Hence, when
coherency exists between InP and GaP, we predict that photo-excited
electrons will be localized in this state, giving rise to an unusual
dependence of the band gap on size.  This effect is peculiar to
coherently strained systems, and is absent in free-standing
(colloidal) dots.

\begin{figure}
\centerline{\epsfxsize=7cm \epsfbox{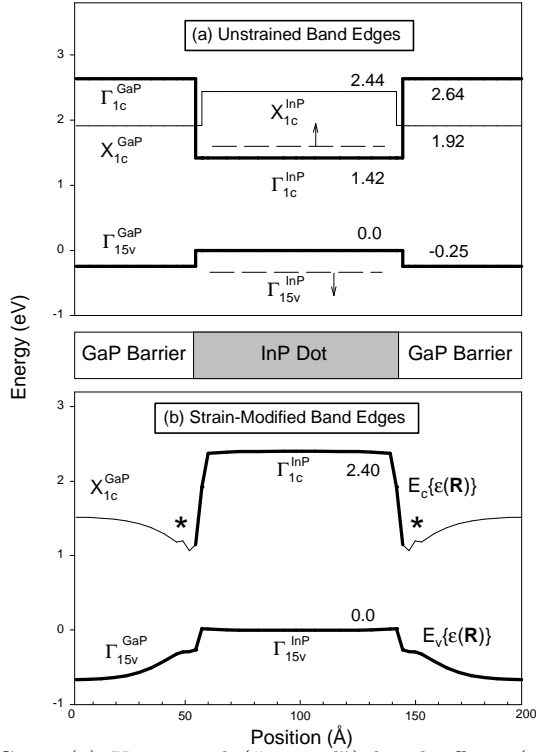}}
\caption{(a) Unstrained (``natural'') band offsets (in eV) between bulk GaP
and InP.  Solid lines indicate bulk band edges and dashed lines
indicate quantum confined levels.  Arrows show the energy change due
to confinement.  $\Gamma$-derived states are shown with thick lines
and $X$-derived states with thin lines.  (b) Strain modified band
edges, $E_{nk}[\epsilon({\bf R})]$, plotted along [100] through the
center of the InP dot with diameter 131\AA~and dot-dot separation of
109\AA.  The lowest(highest) conduction(valence) band is shown at each
position ${\bf R}$.  The * denotes the position at which the lowest
conduction state is localized.}
\label{OFFSETfigure}
\end{figure}

\noindent{\em Expected trends based on band offsets}: Before
presenting our calculated results for GaP-embedded InP dots, we
discuss the basic expectations regarding the nature of the confined
states.  Figure~\ref{OFFSETfigure}a shows our fitted\cite{epm_paper}
{\em unstrained} (``natural'') valence and conduction band offsets
between bulk InP and bulk GaP.  The $\Gamma$ states are shown as heavy
solid lines and the $X$ states as thin solid lines.  We see that InP
can act as a ``well'' both for the conduction band $\Gamma_{1c}$
electrons and the valence band $\Gamma_{15v}$ holes (a ``type I''
offset).  The confined levels are denoted schematically by dashed
lines.  As the InP dot becomes smaller, quantum confinement causes the
confined InP valence levels to be pushed down in energy and the
confined conduction levels to be pushed up (see arrows).  This causes
the $\Gamma_{1c}$-level to be pushed up in energy with respect to the
$X_{1c}$-level in the GaP barrier causing a direct-to-indirect
transition (see below).
In reality, the large (7\%) atomic size-mismatch between GaP and InP
will cause atoms to be displaced off their ideal zincblende positions.
This will alter the effective band offsets of Fig~\ref{OFFSETfigure}a
and thus the anticipated confinements.  To calculate the ensuing
strain we place an InP sphere of radius R inside a large GaP cube, and
relax all the atomic positions to their minimum strain energy values,
using the Valence Force Field (VFF) elastic energy
functional\cite{keating}.  We chose to fix the external dimension of
the GaP cube during the relaxation as this most closely resembles the
experimental situation where InP dots are grown\cite{samuelson94} on a
fixed GaP substrate and the dot-dot separation is large enough to
remove any dot-dot interactions.  However, in the systems studied
here, the large barrier sizes create such a high GaP:InP ratio ($\sim$
30:1) that any external relaxation would be minimal in any case. The
resulting strain exhibits nontrivial hydrostatic and biaxial
components.  Our quantum mechanical calculation of the energy levels
of the dot (see below) will include the effect of such a strain
profile.  However, in order to understand these results, we first
consider a simpler case, namely we calculate the band edge states of
{\em bulk} InP and {\em bulk} GaP subject to the local strain,
$\epsilon({\bf R})$, experienced by the GaP embedded InP dot at
position ${\bf R}$.  To do this we discretize the GaP/InP
nanostructure into ``cells'' with position vector {\bf R} and then
perform $\sim$40 bulk band structure calculations of InP and GaP,
using the empirical pseudopotential method\cite{chelikowsky76}, thus
obtaining the bulk eigenvalues $E_{nk}[\epsilon({\bf R})]$ for band
$n$ at wavevector $k$ within each cell.  Each bulk calculation,
$E_{nk}[\epsilon({\bf R})]$, uses the In-P or Ga-P bond geometry
within that cell.  The resulting strain-modified band edge states are
shown in Fig.~\ref{OFFSETfigure}b.  Compared with the unstrained
offsets (Fig.~\ref{OFFSETfigure}a), we see that the GaP $X_{1c}$-band
edge that is flat in the absence of strain (Fig.~\ref{OFFSETfigure}a)
is now transformed into an attractive trough (indicated by * in
Fig.~\ref{OFFSETfigure}b), capable of localizing electrons.  The
formation of this trough is initially surprising as the deformation
potential at the $X_{1c}$-point is positive and one might therefore
expect the hydrostatic expansion of the GaP at the interface with the
InP dot to drive the $X_{1c}$-state up in energy.  However, the above
bulk calculations show that it is the {\em biaxial} strain present at
this interface which is the dominant term, and this is capable of
forming the electron troughs.  The {\em atomistic} strain has
therefore profoundly modified the nature of the confined electron
states from delocalized to localized.  It is important to emphasize
that conventional\cite{jaros96} calculations of strain modified
conduction band offsets include only the hydrostatic (no biaxial) term
and only the $\Gamma_{1c}$ (no $X_{1c}$) conduction band, and would
therefore miss the important changes in the conduction band edges
between Figs.~\ref{OFFSETfigure}a and \ref{OFFSETfigure}b which our
calculations show are due to the effect of biaxial deformation on the
$X_{1c}$ state.

\noindent{\em Results of calculations on dots}: To calculate the
energy levels of GaP embedded InP dots, we again place an InP dot of
radius R, surrounded by sufficiently thick GaP barrier in a
``supercell'', repeated periodically to create a lattice of dots.
Having created (artificial) translational periodicity, band
theoretical methods can then be applied to study the electronic
properties.  The limit of an isolated dot is achieved by increasing
the thickness of the GaP barrier.  The calculations for both the bulk
bands and the quantum dot levels are based on the atomistic
Hamiltonian
\begin{equation}\label{hamiltonian}
\hat{H} = -\frac{1}{2}\nabla^2+\sum_{\alpha,n}v_\alpha({\bf r}-{\bf
R}_{\alpha n})  \;\;\;.
\end{equation}
The total potential is constructed from screened atomic
pseudopotentials, $v_\alpha$, where $\alpha$ represents Ga, In and P,
and ${\bf R}_{\alpha n}$ are the relaxed atomic positions.  The
pseudopotentials, $v_\alpha$, have been fitted\cite{epm_paper} to the
experimental band gaps, deformation potentials and effective masses.
We use the analytic form of the pseudopotential described in
Ref.\cite{epm_paper}, which was designed to build in the effects of
strain experienced by each atom in lattice mismatched systems.

The supercells studied in this paper contain up to one million atoms,
which is too large for the Hamiltonian in Eq.(\ref{hamiltonian}) to be
solved by direct diagonalization.  We thus use the ``Folded Spectrum
Method'' (FSM) \cite{wang94,kamat}, in which one solves for the
eigenstates of the equation
\begin{equation}
\left( \hat{H}-\epsilon_{\mbox{ref}} \right)^2\psi_i
= (\epsilon-\epsilon_{ref})^2\psi_i \;\;\; ,
\end{equation}
where $\epsilon_{\mbox{ref}}$ is a reference energy, and the
wavefunctions, $\psi_i$, are expanded in a plane wave basis.  By
placing $\epsilon_{\mbox{ref}}$ within the gap, and close to the
valence band maximum or conduction band minimum, one is then able to
obtain the top few valence states or the bottom few conduction states
respectively.  Using this approach the computational cost scales as
$MNlog(N)$, where $N$ is the number of desired electronic
states and $M$ is the number of plane wave basis functions ($M \approx
20$ million in the largest system studied here).  The simulations in
this paper were performed using a parallel code on the Cray T3E900 on
up to 256 processors where the ellipsoid of ${\bf g}$-vectors is
divided over the processors in a similar way to the method used by
Clarke {\em et al}\cite{clarke}. Using this data distribution and fast
parallel FFT's, almost linear speed up with the number of processors
can be obtained for the large systems studied.

\begin{figure}
\centerline{\epsfxsize=7cm \epsfbox{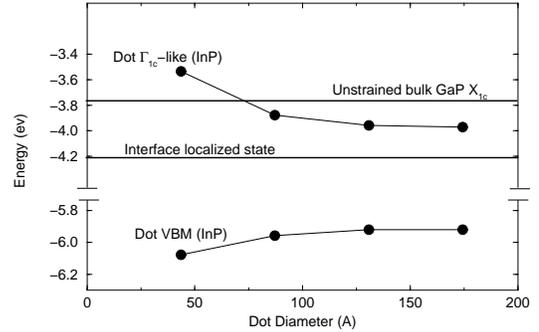}}
\caption{Energies of the near edge states of GaP-embedded InP dots with diameters 44, 87, 131 and 174~\AA~and a dot-dot
separation of 65\AA}
\label{LCBBfigure}
\end{figure}

\begin{figure}
\centerline{\epsfxsize=7cm \epsfbox{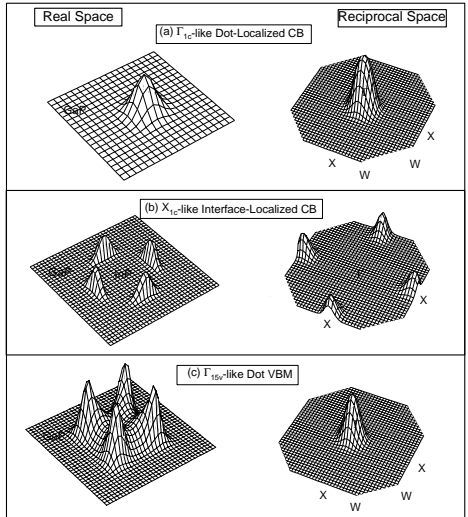}}
\caption{Wavefunction squared (left) and momentum-space analysis
(right) for the near edge states (see Fig.~\ref{LCBBfigure}) of an InP
dot with a diameter of 131\AA~and dot-dot separation of 109\AA. The
left hand side shows each wavefunction squared in the (001) plane
through the center of the InP dot.  The right hand side shows the
momentum-space projection of each wavefunction in the ${\bf k}_z=0$
plane of the Brillouin zone.  Wavefunctions (a) and (c) are
$\Gamma$-derived, and (b) is $X$-derived.}
\label{wavefunctions}
\end{figure}

To facilitate comparison with the spherical free standing InP quantum
dots studied in Refs.\cite{fu97}, we constructed a series of
supercells containing spherical InP quantum dots with diameters of 44,
87, 131 and 174~\AA.  Each dot was surrounded with sufficient GaP
barrier material to produce dot-dot separations of 109, 152, 196 and
239~\AA~respectively.  The calculated energies of the highest occupied
valence states and lowest empty conduction states are shown in
Fig.~\ref{LCBBfigure}.  The left hand side of Fig.~\ref{wavefunctions}
illustrates the corresponding wavefunctions squared of the 131~\AA~
dot.  We see that the highest energy valence wavefunction is localized
within the InP dot (Fig.~\ref{wavefunctions}c), whereas the lowest
conduction wavefunction is localized in pockets at the \{001\} facets
of the interface between the InP dot and the GaP barrier
(Fig.~\ref{wavefunctions}b).  The energy of this interfacial state is
considerably lower than the unstrained bulk GaP $X_{1c}$-state (solid
line in Fig.\ref{LCBBfigure}) for all the dots studied.  To establish
the identity of these wavefunctions in terms of the parent GaP and InP
bulk states, we project the dot wavefunctions, $\psi_i$, into the
zincblende Brillouin zone using the method described in
Ref.\onlinecite{wang97}.  This mapping is shown on the right hand side
of Fig.~\ref{wavefunctions} for the ${\bf k}_z=0$ plane through the
Brillouin zone.  We see that the highest energy valence state is a
$\Gamma$-derived state (Fig.~\ref{wavefunctions}c), while the lowest
conduction state is $X$-derived (Fig.~\ref{wavefunctions}b).  The
calculated dipole transition matrix element between these states is 5
orders of magnitude smaller than one would expect between a more
typical pair of $\Gamma$-derived conduction and valence states,
rendering the transition forbidden.
To investigate the effect of dot-dot separation, a series of similar
calculations were performed on the quantum dot with a diameter of
87~\AA, where the dot-dot separation was steadily increased to values
of 152, 196 and 240~\AA.  The energies of the $X$-derived, interface
localized states were -4.15, -4.22 and -4.22 eV respectively.  The
invariance of this state with dot-dot separation strongly suggests
that the strain induced interface localization evident in
Fig~\ref{wavefunctions}b is not due to too small a choice of
supercell, but an intrinsic effect present at such a interface.
Given that the lowest-energy conduction state, as illustrated in
Fig.~\ref{wavefunctions}b is an $X_{1c}$-derived, interfacial state,
one wonders where is the $\Gamma_{1c}$-derived dot conduction state.
It would not be practical to use the FSM to search for this state as
the large size of the supercell folds many eigenstates to the
$\overline{\Gamma}$-point (at which the FSM calculation is performed)
between the lowest conduction state and the $\Gamma_{1c}$-derived
state.  We instead use the Linear Combination of Bulk Bands
(LCBB)\cite{lcbb} method which specifically searches for states of a
given symmetry (e.g. $\Gamma_{1c}$-derived), whether or not they are
the lowest energy.  Here one first solves for a set of bulk Bloch
wavefunctions, $\phi_{nk}$, of the two materials, InP and GaP.  Then
the Hamiltonian from Eq.(\ref{hamiltonian}) is diagonalized within the
basis of these wavefunctions.  The LCBB method allows one to choose
which bulk Bloch wavefunctions to include in the basis set.  As we are
only interested here in the $\Gamma_{1c}$-derived energy levels, we
include bulk Bloch wavefunctions from a radius of $\frac{14\pi}{l}$
(in reciprocal space) around the $\Gamma$-point, where $l$ is the
supercell length.  The resulting eigenstates were found to be
converged with respect to the basis size at this radius.  The
wavefunction squared of the $\Gamma_{1c}$-derived state for a system
containing a dot with a diameter of 131~\AA~ is plotted in
Fig.~\ref{wavefunctions}a.  The energies of the $\Gamma_{1c}$-derived
states in each of the four InP quantum dots are shown in
Fig.~\ref{LCBBfigure}.  Figure.~\ref{LCBBfigure} shows that there is a
critical dot diameter around 60~\AA~ below which the
$\Gamma_{1c}$-derived conduction state in the InP quantum dot is
higher in energy than the bulk $X_{1c}$-state of the GaP barrier.
This is a type I to type II transition.  However, for all sizes of InP
dot, the $\Gamma_{1c}$-derived state is higher in energy than the
$X_{1c}$-like interfacial state.


In conclusion, we have shown that: (i) the effects of quantum
confinement and pressure raise the energy of the $\Gamma_{1c}$-derived
state in an InP quantum dot so that for dots smaller than 50\AA, this
state is higher than the $X_{1c}$ state of the unstrained GaP barrier.
This transition is analogous to the AlAs-embedded GaAs dot, where the
CBM moves from GaAs-$\Gamma_{1c}$ to AlAs-$X_{1c}$ as the GaAs size
decreases.  However, (ii) strain induces an even lower energy state,
indirect in reciprocal space and localized in real space at the
interface between the InP dot and the GaP barrier.  Therefore, we
predict that, even for large, spherical InP dots, as long as coherency
is maintained, the effects of strain create a system with an indirect
band gap, that is considerably reduced due to the low lying
interfacial state.  This is in direct contrast to the the behavior of
free standing InP dots, which are direct over the large range of
experimentally accessible sizes.

\noindent{\bf Acknowledgements} This work was supported by
United States Department of Energy -- Basic Energy Sciences, Division
of Materials Science under contract No. DE-AC36-83CH10093.  The
calculations were performed using the Cray T3E, located at the
National Energy Research Scientific Computing Center, which is
supported by the Office of Energy Research of the U.S. Department of
Energy.


\end{document}